\documentclass[aps,pra,twocolumn,showpacs,superscriptaddress,bibliography]{revtex4-1}
\usepackage{graphicx}    
\usepackage{dcolumn}    
\usepackage{bm}             
\usepackage{amssymb}   
\usepackage{amsmath}    
\usepackage{subfigure}    
\usepackage{braket}
\usepackage{mathrsfs}
\usepackage{multirow}
\usepackage[export]{adjustbox}
\usepackage[toc,page]{appendix}
\begin{document}

\title{Passive dynamical decoupling of trapped ion qubits and qudits}

\author{R. T. Sutherland}
\email{robert.sutherland@quantinuum.com}
\address{Quantinuum, 303 S Technology Ct, Broomfield, CO 80021, USA}
\address{Department of Electrical and Computer Engineering, University of Texas at San Antonio, San Antonio, Texas 78249, USA}
\author{S. D. Erickson}
\address{Quantinuum, 303 S Technology Ct, Broomfield, CO 80021, USA}

\date{\today}

\begin{abstract}
We propose a method to dynamically decouple every magnetically sensitive hyperfine sublevel of a trapped ion from magnetic field noise, simultaneously, using integrated circuits to adiabatically rotate its local quantization field. These integrated circuits allow passive adjustment of the effective polarization of any external (control or noise) field. By rotating the ion's quantization direction \textit{relative} to this field's polarization, we can perform `passive' dynamical decoupling (PDD),  inverting the linear  Zeeman sensitivity of every hyperfine sublevel. This dynamically decouples the entire ion, rather than just a qubit subspace. Fundamentally, PDD drives the transition $m_{F}\rightarrow -m_{F}$ for every magnetic quantum number $m_{F}$ in the system\textemdash with only one operation\textemdash indicating it applies to qudits with constant overhead in the dimensionality of the qudit. We show how to perform pulsed and continuous PDD, weighing each technique's insensitivity to external magnetic fields versus their sensitivity to diabaticity and control errors. Finally, we show that we can tune the sinusoidal oscillation of the quantization axis to a motional mode of the crystal in order to perform a laser-free two qubit gate that is insensitive to magnetic field noise.
\end{abstract}
\maketitle

\section{Introduction}\label{sec:intro}
Trapped ions offer high fidelity one- and two-qubit gates, long memory times, and the potential to reduce circuit depths with the all-to-all connectivity enabled by ion transport and reordering \cite{harty_2014, ballance_2016, gaebler_2016, srinivas_2021, clark_2021, moses_2023, malinowski_2023}. Regardless, many challenges remain when integrating the capabilities promised in isolated academic demonstrations into larger systems. One reason for this is that large-scale computers must run many distinct operations that, sometimes, have conflicting requirements. For example, many two qubit gating schemes \cite{mintert_2001,ballance_2016,weidt_2016,sutherland_2019_2,srinivas_2021,clark_2021} require shelving each ion to a magnetic field (B-field from here on) sensitive (Zeeman) qubit before implementation, leaving it vulnerable to memory errors. This is typically ameliorated with a spin-echo (dynamical decoupling) sequence \cite{viola_1998_pra,viola_1999_prl} which exchanges a qubit's states to invert it's B-field sensitivity. If a transition between our choice of qubit states cannot be driven, the need for shelving or dressing pulses will complicate any scheme. Further, since exchanging two states only works on a qubit, extending the scheme to qudits ultimately adds control complexity/errors \cite{ringbauer_2019,hrmo_2023}. \\

\begin{figure}[b]
\includegraphics[width=0.5\textwidth]{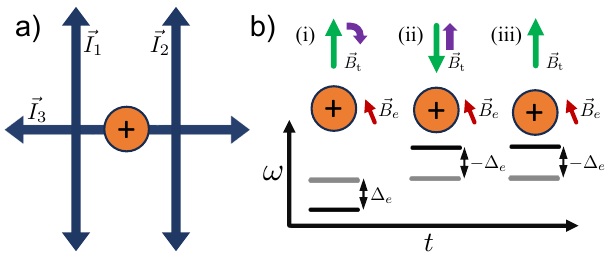}
\centering
\caption{(a) Example circuit design capable of arbitrarily changing the effective quantization field experienced by a `target' ion. (b) Representation of a spin-echo sequence via adiabatic rotation of the local quantization field $\vec{B}_{\text{t}}$ on a target qubit. The ion experiences a quantization field initially pointing up on the page (i), that is then rotated into a direction orthogonal to its original direction, continuing its rotation until it points down (ii). Finally, the system is returned to its initial $\vec{B}_{\text{t}}$ without rotating $\vec{B}_{\text{t}}$. This causes the ion to remain in the state defined by the quantization field in (ii), inducing a $\ket{F,m_{F}}\rightarrow \ket{F,-m_{F}}$ transition for every Zeeman state, i.e. inverting the ion's energy shifts from external B-fields.}
\label{fig:fig_1}
\end{figure}

In this work, we describe a method for dynamical decoupling that inverts the (linear) magnetic sensitivity of every $m_{F}\neq 0$ state of a target ion. Since it affects all states separately, it works equally well on qudits and qubit states that cannot be driven directly. Using trap integrated circuits \cite{ospelkaus_2008,ospelkaus_2011,warring_2013_prl,harty_2014,harty_2016,srinivas_2018, srinivas_2021,malinowski_2023} (see Fig.~\ref{fig:fig_1}a) we can locally manipulate the quantization field direction experienced by a target ion. As we will show, adiabatically inverting the quantization field direction also inverts the B-field sensitivity of the entire ion, letting us increase memory times via dynamical decoupling. Specifically, in Sec.~\ref{sec:theory} we discuss how one can temporarily alter the effective size and direction of a quantization field local to an ion, perform a specific task in the customized environment, then return to the ion's permanent quantization field by ramping the circuits off. Then, in Sec.~\ref{sec:pdd}, we show how the technique can be used to passively dynamically decouple (PDD) the ion from magnetic field noise. By adiabatically rotating the local quantization field until it is anti-parallel to its original direction, we can invert the ion's B-field sensitivity (see Fig.~\ref{fig:fig_1}). In other words, we drive the transition $\ket{F,m_{F}}\rightarrow \ket{F,-m_{F}}$ for every state in the ion. This allows us to dynamically decouple \cite{viola_1998_pra,viola_1999_prl} all internal states of the ion from B-field noise with no need to directly drive a specific transition. The fact that PDD acts on an entire ion, rather than a qubit subspace of that ion, extends dynamical decoupling to qudit systems with constant overhead in the dimensionality of the qudit. In Secs.~\ref{sec:pulsed_pdd} and \ref{sec:cont_pdd}, we discuss how to perform pulsed- and continuous-PDD. Extending from the latter, in Sec.~\ref{sec:gate} we propose a new scheme for laser-free two qubit gates where the rotation frequency of the quantization field is tuned near the motional sideband frequency of a multi-ion crystal in a static magnetic field gradient. The gating scheme promises some of the advantages of those based on oscillating gradients \cite{ospelkaus_2008,ospelkaus_2011,harty_2016,sutherland_2019,sutherland_2019_2,srinivas_2021}, while requiring only a static gradient and remaining insensitive to B-field noise. Finally, in Sec.~\ref{sec:errors} we discuss the impact of diabaticity, cross-talk, and anticipated control errors.

\section{Theory}\label{sec:theory}

We consider a set of `target' ions experiencing two magnetic fields. The first is the permanent quantization field $\vec{B}_{0}=(0,0,B_{0})$, identical for every ion in the computer, and the second is a temporary/local B-field from the near-field of the trap circuits $\vec{B}_{\text{c}}$. We consider the effect these two fields have on the Zeeman states $\ket{F,m_{F}}$ of a system, where $F$ is the total angular momentum of the state, and $m_{F}$ is its magnetic quantum number. This makes the system's `permanent' Hamiltonian: 
\begin{eqnarray}\label{eq:h0_orig}
    \hat{H}_{0}=\frac{\hbar A}{2}\vec{I}\cdot\vec{J} + \mu_{\mathrm{B}}B_{0}(g_{\mathrm{J}}\hat{J}_{\mathrm{z}}+ g_{\mathrm{I}}\hat{I}_{\mathrm{z}}),
\end{eqnarray}
where $A$ is the hyperfine splitting and $\vec{L}\equiv (\hat{L}_{x},\hat{L}_{y},\hat{L}_{z})$ the angular momentum operators $\vec{L}\in \{\vec{I},\vec{J}\}$. While the following results are general, for clarity we consider only $S_{1/2}$ ground state manifolds, meaning the system has two possible values for a state's total angular momentum $F^{+}=I+J$ and $F^{-}=I-J$ \cite{edmonds_1996}. In the following we will write the magnitude of vectors $|\vec{V}|$ as $V$, and their unit vectors $\vec{V}/V$ as $\hat{V}$. We consider only magnetic field magnitudes of $\lesssim 10$ Gauss, as used in current commercial trapped ion processors \cite{pino_2021, moses_2023}. This makes the $\sim~\text{MHz}$ transition Rabi frequencies small relative to the $A/2\pi \sim~\text{GHz}$ frequency separation of typical $S_{1/2}$ hyperfine manifolds \cite{langer_2006}. This allows a perturbative treatment of these off-diagonal elements, resulting in a repeatable AC Zeeman shift. Therefore, we simplify our analysis by making the rotating wave approximation with respect to these terms, i.e. we drop matrix elements between states with different values for $F$. After this approximation, we are free to neglect the $\propto \vec{I}\cdot\vec{J}$ term in Eq.~(\ref{eq:h0_orig}). This reduces Eq.~(\ref{eq:h0_orig}) to:
\begin{eqnarray}\label{eq:h0_simplified}
\hat{H}_{0}\simeq \mu_{\mathrm{B}}B_{0}\Big(g_{\mathrm{J}}\hat{J}_{\mathrm{z}} + g_{\mathrm{I}}\hat{I}_{\text{z}}\Big).
\end{eqnarray}
We explore values of $B_{0}$ to be $\lesssim 10$ Gauss, so that a nearby integrated circuit should be capable of generating a B-field $\vec{B}_{\text{c}}$ larger than $B_{0}$ (with an experimentally feasible current); for example, a wire carrying $0.25~$Amps of current generates $10~$Gauss at a point $50~\mu\text{m}$ away from the trap, much less than the $\sim 1~$Amp surface currents that have were used in recent high-fidelity gate operations \cite{srinivas_2021}. This redefines the total quantization field for the target ion as $\vec{B}_{\text{t}}\equiv \vec{B}_{0}+\vec{B}_{\text{c}}$ for a user-specified duration. Using 1-3 spatially separated circuits integrated into the plane of the trap, we obtain 1-3 degrees of freedom (respectively) to control the magnitude and direction of the total magnetic field $\vec{B}_{\mathrm{t}}=(B_{\text{c},x},~B_{\text{c},y},~B_{\text{c},z}+B_{0})$ experienced by the ion (see Fig.~\ref{fig:fig_1}a). In this work, we will define $\vec{B}_{\mathrm{c}}$ such that it has no projection along the $y$-direction, making $\vec{B}_{\mathrm{t}}=(B_{\text{c},x},~0,~B_{\text{c},z}+B_{0})$. Since $g_{\text{J}}\gg g_{\text{I}}$, it is straightforward to rotate $\vec{B}_{\text{t}}$ at a rate that is slow compared to the electron spin interaction ($\mu_{B}B_{\text{t}}g_{\text{J}}/2\pi\hbar$) but (nearly) instantaneous compared to the nuclear spin interaction ($\mu_{B}B_{\text{t}}g_{\text{I}}/2\pi\hbar$). This means we can drop the $\propto \hat{I}$ term in Eq.~(\ref{eq:h0_simplified}) for simplicity, noting the following scheme does not invert the shift from the nucleus because inverting that shift would require operation timescales $\sim g_{\text{J}}/g_{\text{I}}$ longer than those we discuss here. Considering this, when $\vec{B}_{\text{c}}$ has been ramped on, the total Hamiltonian becomes:
\begin{eqnarray}\label{eq:quantization_orig}
     \hat{H}_{\mathrm{t}} = \mu_{\mathrm{B}}g_{\mathrm{J}}\vec{B}_{\mathrm{t}}\cdot\vec{J},
\end{eqnarray}
which we will use in the numerical examples below. \\

The operator $\hat{U}_{\mathrm{t}}$ that diagonalizes $\hat{H}_{\mathrm{t}}$ can be represented as a rotation of the system such that its quantization field direction is redefined to be along $\hat{B}_{\text{t}}$. This `redefinition' can be encapsulated by a single rotation about an axis orthogonal to both $\vec{B}_{0}$ and $\vec{B}_{\mathrm{t}}$\textemdash here taken to be the $y$-direction. We can represent such a rotation by: 
\begin{eqnarray}\label{eq:rot_op}
    \hat{U}_{\mathrm{t}} = e^{-i\phi\hat{J}_{y}},
\end{eqnarray}
where $\tan(\phi[t])= -B_{\mathrm{t},x}/B_{\mathrm{t},z}$. Using this operator to transform Eq.~(\ref{eq:quantization_orig}) according to $\tilde{H}=\hat{U}^{\dagger}_{\text{t}}\hat{H}_{\text{t}}\hat{U}_{\text{t}}+i\hbar\dot{\hat{U}}^{\dagger}_{\text{t}}\hat{U}_{\text{t}}$ gives:
\begin{eqnarray}\label{eq:h_i}
    \tilde{H}&=& \mu_{\mathrm{B}}B_{\text{t}}g_{\mathrm{J}}\hat{J}_{z} + \hbar\dot{\phi}(t)\hat{J}_{y}.
\end{eqnarray}
In the limit that we change the quantization field slowly compared to the ion's Zeeman splitting, we can let $\dot{\phi}\rightarrow 0$ and ignore the latter term in the above equation\textemdash its largest effect being an added, repeatable AC contribution to the effective value of $B_{\text{t}}$, which can be calibrated out (see Sec.~\ref{sec:errors}).  

\subsection*{Passive Field Rotations}\label{sec:calibrate}

When the ion experiences a second external field after its quantization direction has been rotated, the second field's effective polarization will be defined by its relation to $\vec{B}_{\text{t}}$, not $\vec{B}_{0}$. In other words, if an ion would have experienced an operator $\vec{F} = (\hat{F}_{x},\hat{F}_{y},\hat{F}_{z})$ when the ion's quantization field was $\vec{B}_{0}$, the ion will experience the operator $\tilde{F}\equiv \hat{U}_{\text{t}}^{\dagger}\vec{F}\hat{U}_{\text{t}}$ after we rotate $\vec{B}_{0}\rightarrow\vec{B}_{\text{t}}$. For the rotation described by Eq.~(\ref{eq:rot_op}), this would give:
\begin{eqnarray}\label{eq:rotated_f}
    \tilde{F}=(\hat{F}_{x}\cos[\phi]+\hat{F}_{z}\sin[\phi], \hat{F}_{y}, \hat{F}_{z}\cos[\phi]-\hat{F}_{y}\sin[\phi]).
\end{eqnarray}
In a sense, this ``passively" rotates the polarization of the external field relative to the target ion\textemdash independent of our ability to change its lab-frame polarization. It is often difficult to actively change the polarization of a control field, so the ability to control this polarization electronically could simplify many experiments with conflicting polarization requirements. Interestingly, rotating $\hat{B}_{\text{t}}$ rotates the effective polarization of stray magnetic fields as well, providing a unique way of mitigating their harm.

\section{Passive Dynamical Decoupling}\label{sec:pdd}

For any ion interacting with a stray magnetic field, its $1^{\text{st}}$-order Zeeman shifts are proportional to the field's projection onto its local quantization direction $\hat{B}_{\text{t}}$. Therefore, if we adiabatically rotate $\hat{B}_{\text{t}}$ into $-\hat{z}$, we invert shift from the field. In other words, we drive a $m_{F}\rightarrow -m_{F}$ transition for \textit{every} Zeeman state in the ion. This is a unique benefit of `passive' dynamical decoupling PDD: there is no requirement whatsoever on the ability to directly drive transitions between a system's information carrying states, or even how many information carrying states there are. \\

Traditional methods for dynamically decoupling a quantum system from magnetic field noise involve inverting the magnetic sensitivity of \textit{only} a qubit subspace, not the whole quantum system; this typically requires driving a transition (directly or indirectly) between the two states. There is no such requirement for PDD, since it affects every Zeeman state in the system. For example, if we perform a gate on $^{137}$Ba$^{+}$ with one qubit state defined as $\ket{S_{1/2}, F, m_{F}}$ and one as $\ket{D_{5/2}, F^{\prime},m_{F}^{\prime}}$, any dynamical decoupling sequence via traditional means would require a laser beam to drive a transition between the $S_{1/2}$ and $D_{5/2}$ manifolds. With passive dynamical decoupling there is no such requirement since the protocol simply maps the qubit onto $\{\ket{S_{1/2}, F, -m_{F}}, \ket{D_{5/2}, F^{\prime},-m_{F}^{\prime}}\}$. The applicability of PDD is more general than this, however, since it is a single operation that dynamically decouples entire atoms, rather than qubit subspaces; while we will focus on qubits below, the control sequences we describe would similarly dynamically decouple qudit systems with constant overhead.

\begin{figure}[b]
\includegraphics[width=0.5\textwidth]{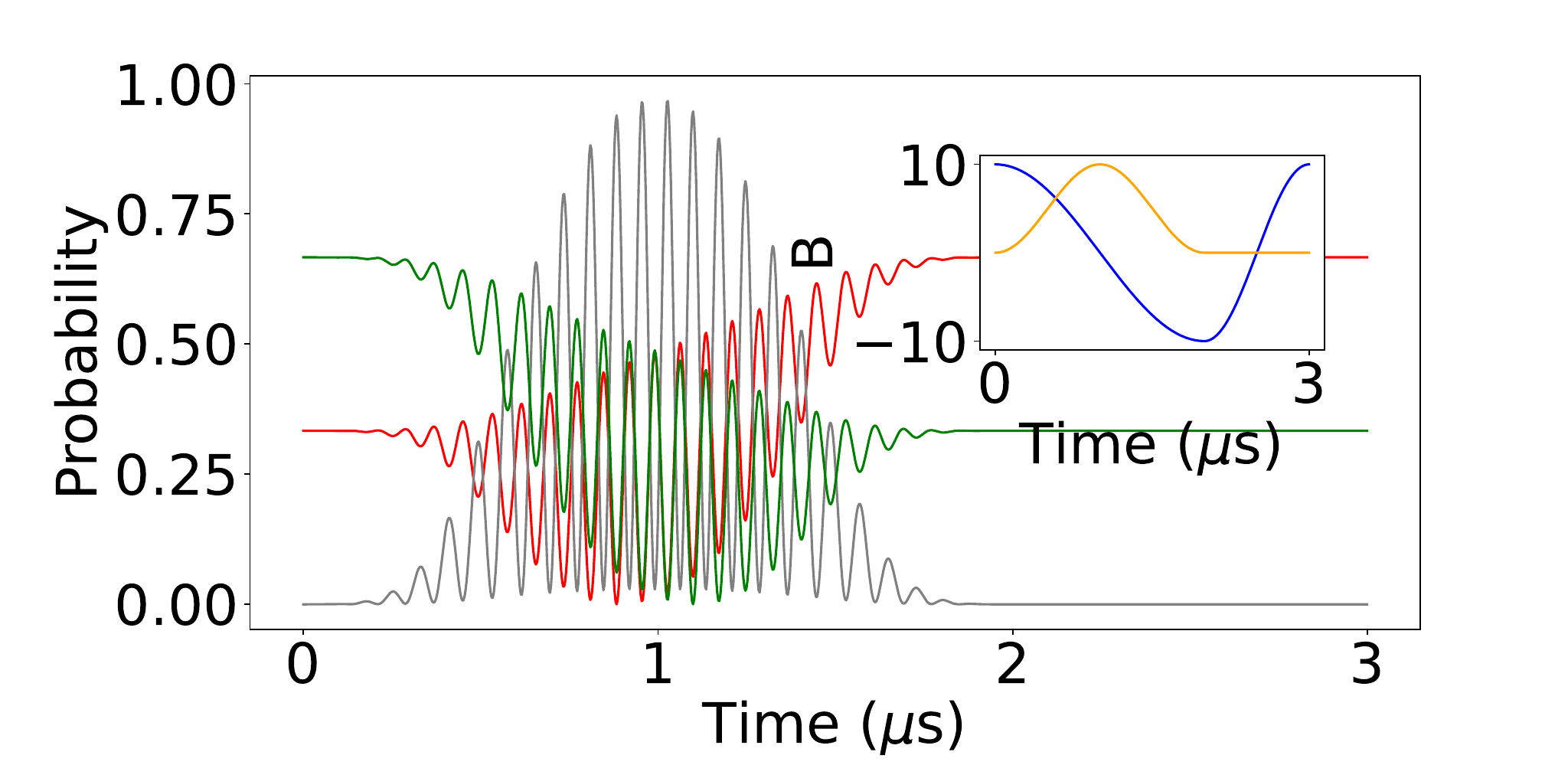}
\centering
\caption{Probabilities versus time of a $^{137}$Ba$^{+}$ ion initialized to $\ket{\psi(0)}= \sqrt{1/3}\ket{F=1,m_{F}=-1} + \sqrt{2/3}\ket{F=1,m_{F}=1}$ of the $S_{1/2}$ manifold of $^{137}$Ba$^{+}$, undergoing an inversion of it's magnetic field sensitivity. Up to a calibratable phase shift, every hyperfine sublevel of the ion undergoes the transformation $m_{F}\rightarrow -m_{F}$ at the end of the sequence. Here, $\ket{F=1,m_{F}=-1}$ is red, $\ket{F=1,m_{F}=0}$ is grey, and $\ket{F=1,m_{F}=1}$ is green. (Inset) The quantization field components experienced by the ion versus time, where $B_{x}=\sin^{2}(\pi t/\tau)$ is the orange line and $B_{z}=1-\cos(\pi t /\tau)$ is the blue line.}
\label{fig:fig_2}
\end{figure}

\subsection{Pulsed PDD}\label{sec:pulsed_pdd}
After we rotate $\hat{B}_{\text{t}}$ into $-\hat{z}$, we can return $\hat{B}_{\text{t}}$ to $+\hat{z}$ without reversing the first operation via ensuring the field remains aligned with $\hat{z}$ during the return; if we ensure $|\hat{B}_{\text{t}}\cdot \hat{z}|=1$, Eq.~(\ref{eq:quantization_orig}) will remain diagonal and the system will not return to its original state. The final states of the ion will have undergone an $m_{F}\rightarrow -m_{F}$ transition (see Fig.~\ref{fig:fig_1}). Writing down the Hamiltonian for an ion experiencing an extraneous B-field oscillating at $\omega_{e}$ we get:
\begin{eqnarray}
    \hat{H}_{e}=\mu_{\text{B}}B_{e}\cos(\omega_{e} t)g_{\text{J}}\hat{J}_{z},
\end{eqnarray}
the effect of which we can analyze using Eq.~(\ref{eq:rotated_f}), inserting $\vec{J}$ for $\vec{F}$. When setting $\phi=\pi$, the scheme inverts the shift from the extraneous field on \textit{every} Zeeman state in the target ion, transforming $\hat{H}_{e}$ into $-\hat{H}_{e}$. If needed, this operation could be repeated in a pattern to perform higher-order pulsed PDD sequences \cite{hayes_2012}. Importantly, this does not dynamically decouple the quadratic shift due to $B_{0}$ mixing the two hyperfine manifolds; this means, for example, that pulsed PDD could not be used to increase the memory time of the $\{\ket{F^{+},0},\ket{F^{-},0}\}$ `clock' qubit. \\

\noindent We provide a numerical example of the dynamics of such a transition in Fig.~\ref{fig:fig_2}. Here we show a spin-echo sequence for a system initialized to $\ket{\psi(0)}= \sqrt{1/3}\ket{F=1,m_{F}=-1} + \sqrt{2/3}\ket{F=1,m_{F}=1}$ of the $S_{1/2}$ ground-state manifold of $^{137}$Ba$^{+}$, giving our qubit a B-field sensitivity of $\frac{\partial\omega_{q}}{\partial B} \simeq 2\pi\times 1.4~\text{MHz/Gauss}$. As shown in the inset, we set the time dependence of the temporary quantization field to be $\vec{B}_{\text{t}}= |{B}_{\text{t}}|(1-\cos[\pi t/\tau], 0, \sin^{2}[\pi t/\tau])$, where $\tau=2~\mu\text{s}$ and $B_{\text{t}}=10$~Gauss (see inset of Fig.~\ref{fig:fig_2}. As will be discussed in Sec.~\ref{sec:errors}, choosing functions for $\vec{B}_{\text{t}}$ such that $\vec{B}_{\text{t}}(0)=\vec{B}_{\text{t}}(t_{f})$ and $\dot{\vec{B}}_{\text{t}}(0)=\dot{\vec{B}}_{\text{t}}(t_{f})=0$ significantly reduces the value of $B_{\text{t}}$ needed to reduce state leakage to a given degree. After the rotation, we see that the system has undergone the desired $m_{F}\rightarrow -m_{F}$ transition discussed above. Importantly, while we are driving a transition between $\ket{F=1,m_{F}=1}$ and $\ket{F=1,m_{F}=-1}$, we are not directly coupling them, which would violate selection rules. In the figure we can see this in the fact that $\ket{F=1,m_{F}=0}$ acts as a bus during the operation, being populated only transiently. \\

\begin{figure}[b]
\includegraphics[width=0.5\textwidth]{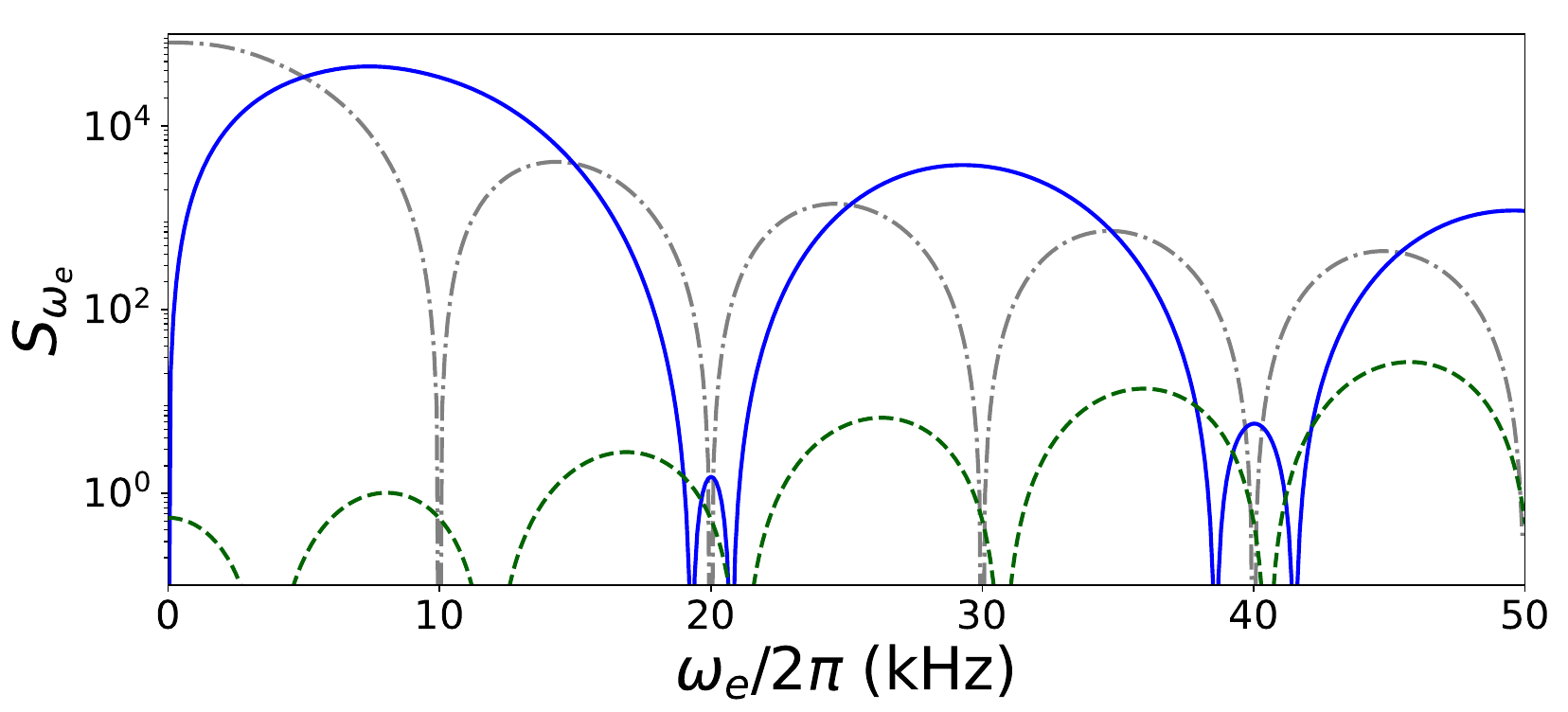}
\centering
\caption{Filter function $S_{\omega_{e}}$ corresponding to the memory error for a $^{137}$Ba$^{+}$ ion initialized to the $\ket{\psi(0)}=\sqrt{2/3}\ket{F=1,m_{F}=1}+\sqrt{1/3}\ket{F=1,m_{F}=-1}$ state of the $S_{1/2}$ ground state manifold experiencing B-field noise for $t_{f}=100~\mu\text{s}$. Each rotation assumes the magnitude of the rotated quantization field to be $B_{\text{t}}= 10$ Gauss. The grey (dashed dotted) line corresponds to an ion undergoing no dynamical decoupling, the blue (solid) line corresponds to an ion undergoing a `passive' spin-echo rotation (pulsed PDD) lasting $\tau=10~\mu\text{s}$ centered at $t_{f}/2$. The green (dashed) line corresponds the scenario where we continuously rotate the quantization field in the $xz$-plane at a rate $\omega_{\text{r}}=2\pi/\tau$ (continuous PDD).}
\label{fig:fig_filter}
\end{figure}

\noindent In Fig.~\ref{fig:fig_filter} we show the increased resistance to B-field noise that results from a `passive' spin-echo, using the same initial state shown in Fig.~\ref{fig:fig_2}; the control sequence is the same, except we set the rotation time to be $\tau=10~\mu\text{s}$. There we plot the filter function $F(\omega)$, defined as the function that, when integrated against the noise spectral density $S(\omega)$, gives the total infidelity of the system $\mathcal{I}=\int^{\infty}_{0}d\omega^{\prime}F(\omega^{\prime})S(\omega^{\prime}) =1-|\braket{T|\psi(0)}|^{2}$ where $\ket{T}$ is the target state. In this example we calculate the memory error of an ion after $t_{f}=100~\mu\text{s}$, so $\ket{T}\equiv \ket{\psi(0)}$ up to a repeatable AC Zeeman phase shift that can be calibrated out (see Sec.~\ref{sec:errors}) of the error budget. We determine $F(\omega)$ numerically by calculating the infidelity of a system experiencing a small B-field, representing a noise term:
\begin{eqnarray}
    \hat{H}_{e} = \mu_{\text{B}}B_{e}\cos(\omega_{e} t + \theta)g_{\text{J}}\hat{J}_{z},
\end{eqnarray}
where we ensure that $B_{e}$ is small enough that $\mathcal{I}\propto B_{e}^{2}$, but large compared to the infidelity due to state leakage. Averaging over $\theta$ and dividing the result by $B_{e}^{2}$ gives $S(\omega_{e})$ for the operation. Comparing $S(\omega_{e})$ for the ion with and without a spin-echo rotation, we can see as $\omega_{e}\rightarrow 0$ that $S(\omega_{e})\rightarrow 0$ for the echoed sequence, while, for the sequence with no PDD, $S({\omega_{e}})$ approaches maximum value as $\omega_{e}\rightarrow 0$; since magnetic field NSDs are typically largest at low frequencies, Fig.~\ref{fig:fig_filter} shows that passive PDD should lead to significantly longer qubit and qudit memory times relative to their non-echoed parallels \cite{sepiol_2019}.

\subsection{Continuous PDD}\label{sec:cont_pdd}

We can perform continuous PDD by rotating the direction of $\hat{B}_{\text{t}}$ about a vector in the $xy$-plane, here taken to be $\hat{y}$. If we keep the value of $B_{\text{t}}$ constant, this gives:
\begin{eqnarray}
    \vec{B}_{\text{t}}=B_{\text{t}}(\sin[\omega_{r}t],0,\cos[\omega_{r}t]),
\end{eqnarray}
rotating $\hat{B}_{\text{t}}$ in a circle at a rate $\omega_{r}\equiv 2\pi/\tau$. As a result, projections of any extraneous B-field $\vec{B}_{e}\cdot\hat{B}_{\text{t}}$ will sinusoidally oscillate at $\omega_{r}$. In other words:
\begin{eqnarray}
    \tilde{H}_{e} \!=\! \mu_{\text{B}}B_{e}\!\cos(\omega_{e} t \!+\! \theta)g_{\text{J}}\Big(\!\hat{J}_{x}\sin[\omega_{r}t],0,\hat{J}_{z}\cos[\omega_{r}t]\!\Big),
\end{eqnarray}
rendering B-field noise where $\omega_{e}\ll\omega_{r}$ off-resonant. We show this in Fig.~\ref{fig:fig_filter} where we plot $S(\omega_{e})$ for the same system described previously, only undergoing continuous PDD. We can see that, for this value of $\tau=10~\mu\text{s}$, continuous PDD suppresses magnetic field noise by several orders-of-magnitude relative to a spin-echo sequence. In the simulation, we suppress state leakage by adiabatically ramping $\vec{B}_{\text{t},x}$ on/off via setting $\vec{B}_{\text{t},x}=\sin^{2}(\omega_{r}t)$ when $\omega_{r}t<\pi/2$ and $(t_{f}-t)\omega_{r} < \pi/2$, ensuring that $\dot{\vec{B}}_{\text{t}}(0)=\dot{\vec{B}}_{\text{t}}(t_{f})=0$. Although we gain insensitivity to ambient B-field noise, continuous PDD requires leaving the ions exposed to uncertainties in $\vec{B}_{\text{t}}$ for significantly longer than pulsed PDD, since it requires the control fields to be on for all $t_{f}$.

\subsection{PDD-assisted laser-free entangling gate}\label{sec:gate}

In the presence of a static B-field gradient, we can rotate $\hat{B}_{\text{t}}$ at a rate tuned near the frequency of a motional mode, driving a spin-dependent force for every state in the ion. This reduces to the the spin-dependent force typically associated with $\hat{\sigma}_{z}\otimes \hat{\sigma}_{z}$ two-qubit gates \cite{molmer_1999,molmer_2000,leibfried_2002} when applied to a qubit subspace. If an ion is at rest in a magnetic field with a non-trivially large gradient pointing in the $z$-direction, its Hamiltonian can be represented as:
\begin{eqnarray}
    \hat{H}_{zz} = \mu_{\text{B}}\Big(B_{z}+\frac{\partial B_{z}}{\partial z}\hat{z}\Big)g_{\text{J}}\hat{J}_{z}.
\end{eqnarray}
The impact of $B_{z}$ is repeatable and can be calibrated out, so we ignore its effect for simplicity. Projecting the position operator $\hat{z}$ onto a specified motional mode with ladder operators $\hat{a}(\hat{a}^{\dagger})$ and frequency $\omega_{a}$, we can make the rotating wave approximation with respect to the other modes in the system and write down:
\begin{eqnarray}
    \hat{H}_{zz}=\hbar\Omega_{zz}\hat{J}_{z}\Big(\hat{a}^{\dagger}e^{i\omega_{a}t} + \hat{a}e^{-i\omega_{a}t}\Big),
\end{eqnarray}
where $\mu_{\text{B}}g_{\text{J}}B_{z}^{\prime}\beta_{a}\equiv \hbar\Omega_{zz}$, and $\beta_{a}$ is the projection of the $\hat{z}$-operator onto mode $a$. If this system undergoes continuous PDD, i.e. $\hat{B}_{\text{t}}$ is rotated in a circle, the transformed Hamiltonian will be:
\begin{eqnarray}
    \tilde{H}_{zz}&\simeq &\hbar\Omega_{zz}\Big(\hat{a}^{\dagger}e^{i\omega_{a}t} + \hat{a}e^{-i\omega_{a}t}\Big)\Big(\hat{J}_{z}\cos[\omega_{r}t]+\hat{J}_{x}\sin[\omega_{r}t]\Big) \nonumber \\
    &&+\mu_{\mathrm{B}}B_{\text{t}}g_{\mathrm{J}}\hat{J}_{z},
\end{eqnarray}
assuming the diabatic term is negligible. Transforming into the interaction picture with respect to the $\propto B_{\text{t}}$ control field term, we make the rotating wave approximation with respect to the $\propto \hat{J}_{x}$ terms\textemdash assuming that $\omega_{a}\pm\omega_{r}$ is far-detuned from the Zeeman splitting. Finally, if we tune $\omega_{r}\sim\omega_{a}$, we get:
\begin{eqnarray}
    \tilde{H}_{zz}^{\prime} = \frac{\hbar\Omega_{zz}}{2}\Big(\hat{a}^{\dagger}e^{i\delta t} + \hat{a}e^{-i\delta t}\Big)\hat{J}_{z},
\end{eqnarray}
where $\delta \equiv \omega_{a}-\omega_{r}$. Projecting this Hamiltonian onto a qubit subspace, we get the typical Hamiltonian for a $\hat{\sigma}_{z}\otimes \hat{\sigma}_{z}$ entangling gate. \\

A key advantage of this gate scheme is that it is continuously dynamically decoupled from B-field noise in a similar way as the laser-free gates in Refs.~\cite{harty_2016,srinivas_2021}, suppressing qubit dephasing during the gate. Another advantage this scheme has over previous laser-free schemes, however, is that there is (again) no requirement on the ability to directly drive a transition between the qubit states. Importantly, this new scheme allows experimentalists to tune near a motional mode of the system in a \textit{static} gradient\textemdash without the need for $\sim \text{GHz}$ frequency fields to drive the qubit transition, as is the case in Refs.~\cite{mintert_2001,weidt_2016}. This means it could be useful in gates that use permanent gradients.

\section{Sources of error}\label{sec:errors}
We have shown how to use PDD to render a system less sensitive to ambient B-field noise during a given operation time $t_{f}$, but have not discussed potential sources of error. In the limit that we can manipulate the currents in trap integrated circuits with errors much smaller than memory errors from B-field noise, PDD offers a clear advantage. If it is not clear that this is the case we must consider the main sources of error intrinsic to the specific PDD scheme we use. In the following we will examine the (rough) extent to which errors from uncertainties in the control field operations, and from diabaticity, should affect the usefulness of PDD. We note that, while we do not examine crosstalk in detail since this will likely be device specific, idle qubits far way from the source of $\vec{B}_{\text{c}}$ will only experience small repeatable perturbations to $\vec{B}_{0}$, resulting in phase shifts that can be tracked.

\subsection{Diabaticity}\label{sec:leak}

\begin{figure}[b]
\includegraphics[width=0.5\textwidth]{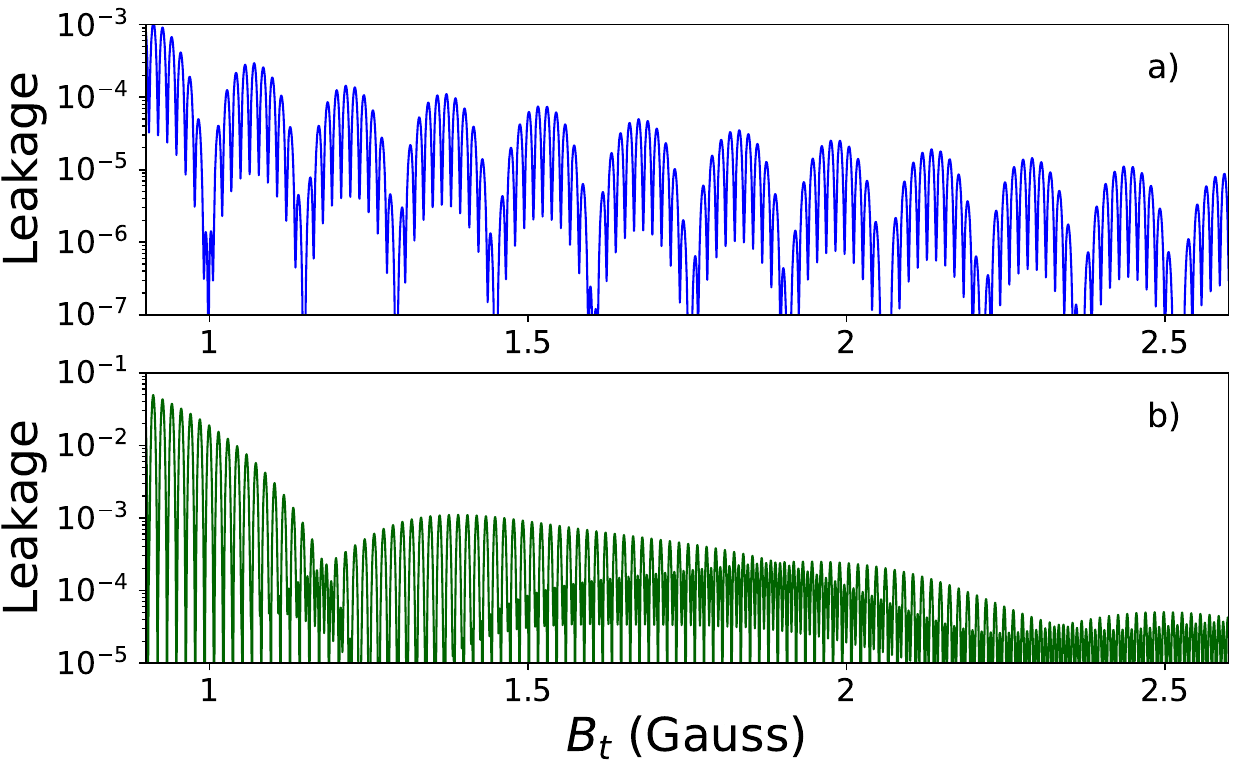}
\centering
\caption{State leakage probability $1-(|\braket{0|\psi(t)}|^2 + |\braket{1|\psi(t)}|^2)$ versus temporary quantization field magnitude $B_{t}$ used to implement dynamical decoupling. For each figure we show pulsed a) and continuous b) dynamical decoupling schemes for a system initialized to the state $\ket{\psi(t)}=\sqrt{2/3}\ket{F=1,m_{F}=1}+\sqrt{1/3}\ket{F=1,m_{F}=-1}$ of the $S_{1/2}$ manifold of $^{137}$Ba$^{+}$. Each set of simulations is for a $t_{f}=100~\mu\text{s}$ sequence.  In a) we show a `passive' spin-echo, where we rotate $\vec{B}_{t}\rightarrow -\vec{B}_{t}$ over $\tau=10~\mu\text{s}$, and in b) we rotate $\vec{B}_{t}$ continuously at $\omega_{r}=2\pi/\tau$.}
\label{fig:fig_leak}
\end{figure}

So far, we have not discussed the role of the $\propto \dot{\phi}(t)$
term in Eq.~(\ref{eq:h_i}), which represents off-resonant diabatic transitions. If we rotate the system slowly enough, it will be fully adiabatic and $\dot{\phi}(t)\rightarrow 0$. Since we cannot guarantee adiabaticity for every system of interest, it is crucial to understand the general behaviour of PDD when $\dot{\phi}(t)\neq 0$. To examine this, we first transform this term into the interaction picture with respect to the $\propto B_{\text{t}}(t)$ quantization field term in Eq.~(\ref{eq:h_i}) using:
\begin{eqnarray}
    \hat{U}_{I} =\exp\Big(-i\varepsilon[t]\hat{J}_{z}\Big),
\end{eqnarray}
where $\varepsilon(t) \equiv(\mu_{\text{B}}g_{\text{J}}/\hbar)\int^{t}_{0}dt^{\prime}B_{\text{t}}(t^{\prime})$ is the time integral of the Zeeman splitting induced by $\vec{B}_{\text{t}}$. This gives:
\begin{eqnarray}\label{eq:h_diab_rot}
    \hat{H}_{\text{t},I} = \hbar\dot{\phi}(t)\Big(\hat{J}_{y}\cos[\varepsilon(t)] + \hat{J}_{x}\sin[\varepsilon(t)] \Big),
\end{eqnarray}
showing the `diabatic' term in Eq.~(\ref{eq:h_i}) can be represented with an angular momentum operator $\vec{J}$ that rotates in the $xy$-plane. We can approximate the time propagator for Eq.~(\ref{eq:h_diab_rot}), using the Magnus expansion \cite{magnus_1954}:
\begin{eqnarray}
    \hat{U} &=&\exp\Big(\frac{-i}{\hbar}\int^{t}_{0}dt^{\prime}\!\hat{H}_{\text{t},I}(t^{\prime}) \\
    &&-\frac{1}{2\hbar^{2}}\int^{t}_{0}\!\!\int^{t^{\prime}}_{0}dt^{\prime}dt^{\prime\prime}\Big[\hat{H}_{\text{t},I}(t^{\prime}),\hat{H}_{\text{t},I}(t^{\prime\prime}) \Big] \Big). \nonumber
\end{eqnarray}
Plugging Eq.~(\ref{eq:h_diab_rot}) into this equation simplifies to:
\begin{eqnarray}\label{eq:diabatic_mag}
    \hat{U}&=&\exp\Big(-i\int^{t}_{0}dt^{\prime}\dot{\phi}(t^{\prime})\Big(\hat{J}_{y}\cos[\varepsilon(t^{\prime})] + \hat{J}_{x}\sin[\varepsilon(t^{\prime})] \Big) \nonumber \\  
    &&-\frac{\hat{J}_{z}}{2}\!\int^{t}_{0}\int^{t^{\prime}}_{0}\!\!\!\!dt^{\prime}dt^{\prime\prime}\dot{\phi}(t^{\prime})\dot{\phi}(t^{\prime\prime}) \sin\Big(\varepsilon[t^{\prime\prime}]-\varepsilon[t^{\prime}], \Big)
\end{eqnarray}
which we can separate into two distinct effects: state leakage, represented by the $1^{\text{st}}$-order terms,
and an additional $\propto \hat{J}_{z}$ shift, represented by the $2^{\text{nd}}$-order terms. If we approximate $\dot{\phi}(t)\rightarrow\phi_{0}$ and $\varepsilon(t)\rightarrow \varepsilon_{0}t$ to be constant over $t_{f}$, we can see that the $1^{\text{st}}$-order terms generally represent off-resonant transitions inducing leakage that scales $\propto (\phi_{0}/\varepsilon_{0})^{2}$. We can further suppress off-resonant leakage errors by ``pulse shaping" $\dot{\phi}(t)$. For our examples of continuous PDD we did this by setting $B_{\text{t},x}=B_{\text{t}}\sin^{2}(\omega_{r}t)$ at the start and end of the control sequence. This ensures $\ddot{\phi}(0)=\ddot{\phi}(t_{f}) = 0$, which further suppresses the probability of an off-resonant transition in the same way it does for carrier \cite{roos_2008,sutherland_2019_2} and spin-motion \cite{sutherland_2023} interactions. In Fig.~\ref{fig:fig_leak} we show increasing the value of $B_{\text{t}}$ also suppresses state leakage, making these transitions more off-resonant by increasing the Zeeman splitting of the atom. We demonstrate this in Fig.~\ref{fig:fig_leak} for the example case used in Figs.~\ref{fig:fig_2} and \ref{fig:fig_filter}, plotting the leakage error probability ($1-|\braket{1,-1|\psi(t_{f})}|^{2}-|\braket{1,1|\psi(t_{f})}|^{2}$) versus $B_{t}$ for both pulsed and continuous PDD. This shows leakage can be suppressed below $10^{-4}$ at $\sim 1$~Gauss for our pulsed and $\sim 2.5$~Gauss for the continuous PDD examples; for reference, a $0.1$ Amp current in an infinitely thin wire produces a $4$~Gauss B-field at distance of $50~\mu\text{m}$. This corresponds to resistive heat loads $\sim 100$ times lower than the $\sim~1$ Amp currents used in Ref~\cite{srinivas_2021}. The $2^{\text{nd}}$-order terms  in Eq.~(\ref{eq:diabatic_mag}) correspond to a $\propto \hat{J}_{z}$ AC shift, that, in general, grows like $\propto \dot{\phi}^{2}/\varepsilon$. If we can consistently repeat both $\dot{\phi}$ and $\varepsilon$ over many operations, then we can track this shift and eliminate its effect.

\subsection{Control field uncertainties}\label{sec:control}

In any experiment, a control field will drift from its `ideal' value, i.e. $\vec{B}_{\text{c}}(t)\rightarrow \vec{B}_{\text{c}}(t)+\delta{\vec{B}_{\text{c}}}(t)$. The biggest source of infidelity from $\delta\vec{B}_{\text{c}}$ will likely be due to its projection onto $\hat{B}_{\text{t}}$, since this will give a $1^{\text{st}}$-order shift that can be described by the time propagator:
\begin{eqnarray}\label{eq:control_error}
    \hat{U}_{\delta B_{\text{c},\text{t}}} \simeq \exp\Big(-\frac{i}{\hbar}\frac{\partial \omega_{q}}{\partial B} \hat{J}_{z}\int^{t_{r}}_{0}dt^{\prime}\hat{B}_{\text{t}}[t^{\prime}]\cdot\delta \vec{B}_{\text{c}}[t^{\prime}]\Big),
\end{eqnarray}
where $t_{r}$ is here the time the control fields are on. Assuming $\delta \vec{B}_{\text{c}}$ drifts slowly over a calibration cycle, we can take $\delta \vec{B}_{\text{c}}$ to be constant and write down an approximate equation for $\mathcal{I}$. We can Taylor expand Eq.~(\ref{eq:control_error}) if $\mathcal{I}\ll 1$, letting us obtain:
\begin{eqnarray}
    \mathcal{I}=\Big(\delta B_{\text{c}}\frac{\partial \omega_{q}}{\partial B}t_{r}\Big)^{2}\lambda^{2}_{\sigma_{z}},
\end{eqnarray}
where $\lambda_{\hat{\sigma}_{z}}^{2}\equiv \braket{\hat{\sigma}_{z}^{2}}-\braket{\hat{\sigma}_{z}}^{2}$ is the variance of the $\hat{\sigma}_{z}$ Pauli-z operator of the qubit subsystem. We here take $\lambda_{\hat{\sigma}_{z}}^{2}=1/3$ upon averaging over SU(2). If we choose $B_{\text{c}}\simeq 2.5~\text{Gauss}$ and assume an uncertainty of $10^{-4}$ for $\delta\vec{B}_{\text{c}}\cdot \hat{B}_{\text{t}}$, we get an infidelity of $\mathcal{I}\simeq 1\times 10^{-5}$ for the pulsed spin-echo case (using $t_{r}\simeq 3~\mu\text{s}$) and $\mathcal{I}\simeq 2\times 10^{-2}$ for continuous PDD (using $t_{r}\simeq 100~\mu\text{s}$). We can see continuous PDD is significantly more sensitive to control field uncertainties because $t_{r}$ is larger, making pulsed PDD more appealing as a near-term tool.

\section{Conclusion}\label{sec:conclusion}

In this work, we discussed how trap integrated circuits can be used to implement arbitrary passive rotations of the quantization axis temporarily experienced by a target ion. This lets us implement `passive' dynamical decoupling by inverting the ion's quantization field's projection onto external B-fields, inverting the magnetic susceptibility of its hyperfine sublevels. We showed how to do `pulsed' PDD, where the energy dependence of every Zeeman state in the ion is reversed over a short timescale, and how to do `continuous' PDD, where the ion's quantization direction is rotated sinusoidally. We also proposed a new way to perform laser-free two-qubit gates by showing we can tune this continuous rotation to the frequency of a motional mode sideband while the ion is in a static gradient. Finally, we discussed potential sources of error for the scheme.

\section*{Acknowledgements}

We would like to thank M. Foss-Feig, J. Gaebler, B. J. Bjork, C. Langer, M. Schecter, C. N. Gilbreth, J. Bartolotta, E. Hudson and P. J. Lee for helpful discussions.

\bibliography{biblio}

\end{document}